\documentclass[structabstract]{aa}  
\usepackage{caption}
\usepackage{graphicx}
\usepackage{epsfig}
\usepackage{amsmath}
\usepackage{rotating}
\usepackage{natbib}
\usepackage{deluxetable}
\usepackage{lscape}
\usepackage[T1]{fontenc}
\usepackage[british]{babel}
\usepackage[varg]{txfonts}
\usepackage{graphicx}
\usepackage{txfonts}
\def \aj {AJ}
\def \apj {ApJ}
\def \apjs {ApJS}
\def \apjl {ApJ}

\def \nat {Nature}
\def \mnras {MNRAS}
\def \aap {A\&A}
\def \aaps {A\&AS}
\def \araa {ARA\&A}
\def \pasp {PASP}

\begin{document}
   \title{Low-frequency High-resolution Radio Observations of the TeV-emitting Blazar SHBL\,J001355.9$-$185406}

   \author{Natalia {\.Z}ywucka 
          \inst{1}
          \and Arti Goyal\inst{1} \and Marek Jamrozy\inst{1} \and Micha{\l} Ostrowski\inst{1} \and {\L}ukasz Stawarz\inst{2,1} 
          }

   \institute{Astronomical Observatory, Jagiellonian University, ul. Orla 171, 30-244 Krak{\'o}w, Poland
              (\email{natti@oa.uj.edu.pl})
   \and Institute of Space and Astronautical Science JAXA, 3-1-1 Yoshinodai, Chuo-ku, Sagamihara, Kanagawa 252-5210, Japan
             }

   \date{Received \today; accepted \today}

 
  \abstract
   { 
   In the framework of the unification scheme of radio-loud active galactic nuclei, 
   BL Lac objects and quasars are the beamed end-on counterparts of low-power (FR\,I) and 
   high-power (FR\,II) radio galaxies, respectively.    
   }
   { 
   Some BL Lacs have been found to possess the FR\,II-type large-scale
   radio morphology, suggesting that the parent population of BL Lacs is a mixture 
   of low- and high-power radio galaxies. This seems to apply only to 
   `low frequency-peaked'  BL Lacs, since all the `high frequency-peaked' BL Lacs
   studied so far were shown to host exclusively the FR\,I-type radio jets. 
   While analyzing the {\emph {NVSS}} survey maps of the TeV detected 
   BL Lacs, we have however discovered that the high frequency-peaked object 
   SHBL\,J001355.9$-$185406 is associated uniquely with the one-sided, 
   arcmin-scale, and edge-brightened jet/lobe-like feature extending to the south-west 
   from the blazar core. 
   }
   { 
 In order to investigate in detail the large-scale morphology of SHBL\,J001355.9$-$185406,
 we have performed low-frequency and high-resolution observations of the source at 156, 
 259 and 629\,MHz using the Giant Metrewave Radio Telescope.
   }
   {
  Our analysis indicates that no diffuse arcmin-scale emission is present around the unresolved 
 blazar core, and that the arcmin-scale structure seen on the {\emph {NVSS}} map breaks into 
 three distinct features unrelated to the blazar, but instead associated with background AGN.
    }
   {
 The upper limits for the extended radio halo around the TeV-emitting BL Lac object
 SHBL\,J001355.9$-$185406 read as $< 10\%-1\%$ at $156-629$\,MHz.
 The fact that the integrated radio spectrum of the unresolved blazar core is flat down to 
 156\,MHz indicates that a self-similar character of the jet in the source 
  holds up to relatively large distances from the jet base.
   }

\titlerunning{Low-frequency Radio Observations of SHBL\,J001355.9$-$185406}
\authorrunning{Natalia {\.Z}ywucka et al. }
   \maketitle
%

\section{Introduction} 
\label{intro}

Blazars constitute the most extreme class of active galactic nuclei (AGN), 
for which the total radiative output is dominated by a Doppler-boosted 
and highly variable emission of a nuclear relativistic jet observed 
at a small viewing angle. The blazar family includes flat-spectrum radio-loud 
quasars (FSRQs) and BL Lacertae (BL Lac) objects. In the framework of the 
``unified scheme'' of AGN, powerful radio galaxies of the Fanaroff-Riley morphological 
type II (FR\,IIs; see, \citealt{1974MNRAS.167P..31Fanaroff}) constitute the parent 
population of FSRQs, while low-power radio galaxies of the Fanaroff-Riley morphological 
type I (FR\,Is) are considered to be ``misaligned'' BL Lacs (\citealt{1989ApJ...336..606Barthel}; 
\citealt{1995PASP..107..803Urry}; \citealt{2009ApJ...694L.107Xu}). 

High-dynamic range radio imaging of BL Lac objects at GHz frequencies often reveals the presence of 
a diffuse extended radio emission, with the integrated luminosity exceeding, in several cases, 
the FR\,I/FR\,II divide. We note that this borderline scales with the optical luminosity of the 
host as $\propto L_{\rm opt}^{1.8}$, with the normalization $L_{\rm 1.4\,GHz} = 10^{24}$\,W\,Hz$^{-1}$ 
for the absolute isophotal magnitude of the host galaxy $-21$ measured to 24.5 magnitudes 
per arcsec$^2$ in the rest-frame of a source (\citealt{1996AJ....112....9Ledlow}). This may 
indicate that the parent population of BL Lac objects is a mixture of FR\,I and FR\,II type galaxies 
(see, e.g., \citealt{1999A&AS..139..601Cassaro}; \citealt{2001AJ....122..565Rector}; 
\citealt{2008MNRAS.391..967Landt}; \citealt{2010ApJ...710..764Kharb}). Therefore, a careful 
investigation of the extended diffuse radio emission in BL Lacs is of importance for the 
models relating the AGN accretion power and jet luminosity in general, and the blazar 
population studies in particular (e.g., \citealt{1982MNRAS.199..883Blandford}; 
\citealt{1991Natur.349..138Rawlings}; \citealt{2011MNRAS.414.2674Ghisellini}).

The high frequency-peaked BL Lacs (HBLs), i.e. blazars with synchrotron peak 
frequencies $> 10^{15}$\,Hz (see the discussion in \citealt{2010ApJ...716...30Abdo}
and references therein), dominate the population of extragalactic TeV 
emitters\footnote{\texttt{http://tevcat.uchicago.edu/}}. These sources, unlike the 
low frequency-peaked BL Lacs (LBLs; synchrotron peak frequencies $<10^{15}$\,Hz), 
seem to be associated strictly with FR\,I-type large-scale radio structures
(e.g., \citealt{2010ApJ...710..764Kharb}), and to form a relatively distinct sub-class
of blazars with respect to the jet physical properties (e.g., \citealt{2010MNRAS.401.1570Tavecchio}).
Also, no superluminal velocities have been detected in the TeV-emitting HBLs on 
milli-arcsec scales till now, despite several dedicated observational programs 
(e.g., \citealt{2010ApJ...723.1150Piner} and references therein). 
Such apparent superluminal velocities are typical for FSRQs and not rare for LBLs.  

SHBL\,J001355.9$-$185406 (J2000.0 R.A.\,$\rm=00^{h}13^{m}56\fs054$, 
Dec.\,$\rm=-18\degr54\arcmin06\farcs48$; redshift $z = 0.0948$ according to 
\citealt{2009MNRAS.399..683Jones}) is a peculiar example of an HBL. 
This source was listed in a catalog of extremely high X-ray--to--radio flux ratio targets 
in the multi-frequency `Sedentary Survey' of HBLs performed 
by \citet{2005A&A...434..385Giommi}. Its $0.1-2.4$\,keV X-ray flux is 
$1.26\times10^{-11}$\,erg\,s$^{-1}$\,cm$^{-2}$, and the 
corresponding $1400$\,MHz radio flux spectral density is $29.2\pm1.0$\,mJy 
(see, \citealt{1998AJ....115.1693Condon}). Very recently, this source has been 
detected at the level of $\sim 1\%$ of the flux of the 
Crab nebula above 300\,GeV by the High Energy Stereoscopic System 
(H.E.S.S.; \citealt{2013A&A...554A..72Hess}). The analysis of the 
{\it Fermi} Large Area Telescope ({\it Fermi}-LAT) data revealed 
a faint and flat-spectrum counterpart of SHBL\,J001355.9$-$185406 
at the GeV photon energies, with the integrated photon flux 
above $100$\,MeV of $(0.9\pm0.7)\times10^{9}$\,ph\,cm$^{-2}$\,s$^{-1}$ 
and the photon index of $1.5\pm0.2$ (\citealt{2010ATel.3014....1Sanchez}). 
At lower frequencies, the discussed object was also reported as a 
bright and variable NIR/optical emitter. In particular, the 
NIR flux of SHBL\,J001355.9$-$185406 determined in 2010 was 
about 0.7\,mag brighter than that previously published in the 
{\emph {2MASS}} survey (\citealt{2010ATel.3023....1Cassarco}). 
The corresponding $70\%$ flux increase imply the jet 
origin of the NIR continuum.

\begin{figure}[t]
\includegraphics[height=8.0cm,width=8.0cm]{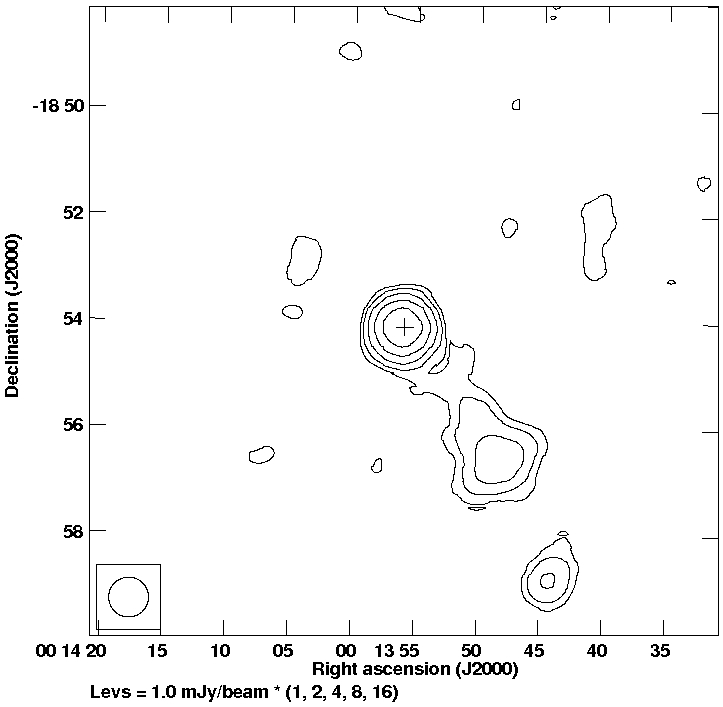}
\caption{ 
Contour plot of the 1400\,MHz {\emph {NVSS}} radio continuum emission.
The plus sign marks the position of blazar core and the bottom left 
inset shows the size of restoring beam.
}
\label{fig.nvss}
\end{figure}

The peculiarity of the SHBL\,J001355.9$-$185406 is related however 
to its large-scale radio structure which have been found when 
examining the archival radio National Radio Astronomy Observatory 
({\emph {NRAO}}) Very Large Array ({\emph {VLA}}) Sky Survey 
({\emph {NVSS}}; \citealt{1998AJ....115.1693Condon}) maps of the TeV-detected BL Lacs. 
This structure, shown in Figure\,\ref{fig.nvss}, consists of a compact core component, 
as expected for a blazar, but also of a one-sided and elongated 
lobe/jet-like feature extending to the south-west from the nucleus.
At the redshift of the source (luminosity distance of $d_{L} = 430$\,Mpc 
for the flat cosmology with $H_{0} = 71$ km sec$^{-1}$ Mpc$^{-1}$, 
$\Omega_{\rm M} =0.27$ and $\Omega_{\Lambda} =0.73$), the 3\farcm3 
extension of the lobe corresponds to the physical (projected) size of $\sim 344$\,kpc 
and the $1400$\,MHz flux $15.8$\,mJy corresponds to the lobe's monochromatic 
luminosity of $\sim 5 \times 10^{39}$\,erg\,s$^{-1}$. The derived extended 
luminosity is therefore comfortably low as for the low-power blazar of 
the HBL type, yet the morphology of the giant lobe is striking 
for two reasons. 

First, the lobe is clearly edge-brightened, and hence the 
whole large-scale radio structure of SHBL\,J001355.9$-$185406 
resembles more an FR\,II-type object.
Second, the extended 
structure in the discussed source is clearly one-sided. This is \emph{not} 
expected for the large-scale structures in radio-loud AGN in general, 
regardless on their Fanaroff-Riley type, since the lobes' expansion velocities 
on scales of tens and hundreds of kpc are expected to be sub-relativistic 
only. In other words, the Doppler (and time-travel) effects cannot be easily blamed 
for the apparent lack of the counter-lobe in SHBL\,J001355.9$-$185406
(but see in this context \citealt{2004ApJ...613..119Stawarz} and \citealt{2007ApJ...670L..85Bagchi}). 
In order to investigate this issue in more detail, we performed deep and high-resolution 
radio observations of SHBL\,J001355.9$-$185406 at low frequencies. The
results of the analysis are presented below. 

\begin{figure}[t]
\includegraphics[height=6.5cm,width=6.5cm]{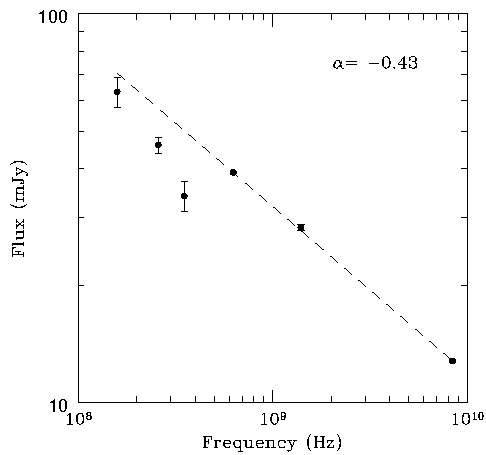}
\caption{
Integrated radio spectrum of the unresolved core in SHBL\,J001355.9$-$185406.
The dashed line shows the result of weighted linear regression analysis. 
}
\label{spec_radio}
\end{figure}

\section{Observations and Data Analysis}
\label{observation}

Observations were carried out using the Giant Metrewave 
Radio Telescope ({\emph {GMRT}}) in three frequency bands, namely 156, 259
and 629\,MHz, the latter two being taken in dual frequency mode 
(\citealt{1991ASPC...19..376Swarup}).
Data were recorded with an integration time of 8 seconds 
with the available frequency band divided
into 256 channels, following the usual protocol of
observing flux and phase calibrator interlaced with observation on
the source. At the beginning and end of the observing run, 
flux calibrator 3C\,48 was observed for about
15 min. The phase calibrator J0025$-$260, was observed 
every $\sim$25 min for $\sim$4-5 min. 
Table\,\ref{obs} gives the summary of the observations.
The data reduction was carried out following standard 
calibration and reduction procedures in the 
Astronomical Image Processing System ({\emph {AIPS}}). 
Data were edited for strong radio frequency interference (RFI) 
and then standard flux and phase calibration were applied to the source.
In addition to the newly obtained {\emph {GMRT}} data, we have also 
analyzed the archival 8434\,MHz {\emph {VLA}} data for 
SHBL\,J001355.9$-$185406 in a standard fashion as described below. 

\begin{table*}[t]
\tiny
\caption{Observational summary }
\label{obs}
\begin{tabular}{cccccccccccc}\\\hline
$\nu_{RF}$ & BW     & Date     &  Time    & Pol. & Flux cal.   & Phase cal.   & HPBW  & Synth. beam  & PA &   $\sigma_{map}$ & Dyn. range \\
(MHz)     &  (MHz)         & (dd.mm.yy) &  (hr)          &   & name(flux) & name(flux)  & (arcmin)   &  (arcsec$^2$)   & (deg)   &   (mJy/b) &  (peak/rms err.) \\
(1)       &  (2)     & (3)    &  (4)           &    (5)   &        (6)   &(7)  & (8) & (9)   & (10) & (11) & (12) \\ \hline
156       &  16      & 29.08.11 & $\sim$ 6   &  RR,LL & 3C 48(65)        & J0025$-$260(20) & 186    & 20.3\,$\times$\,13.0&  23.3       & 2.0    &  $\sim$ 31   \\
259$^*$   &  6.5     & 01.09.11 & $\sim$ 6   &  LL    & 3C 48(47)        & J0025$-$260(18) & 116    & 15.1\,$\times$\,10.5&  37.4       & 0.09    &  $\sim$ 60  \\
629$^*$   &  32      & 01.09.11 & $\sim$ 6   &  RR    & 3C 48(30)        & J0025$-$260(14) & 45    &  6.4\,$\times$\,4.1 &  32.9        & 0.07    &  $\sim$ 600  \\ 
8434$^\dag$&  50      & 06.06.98 & $\sim$ 0.08   &  RR,LL    & 3C 48(3.3)        & J0110$-$076(0.55) & 6    &  0.48\,$\times$\,0.28 &  $-$31.07        & 0.09    &  $\sim$ 150  \\\hline 
\end{tabular}
\begin{minipage}{\textwidth}
Columns: (1) central radio frequency $\nu_{RF}$ of observation 
($^*$ indicates a dual frequency mode; $^\dag$ archival {\emph {VLA}} data 
obtained in AnB configuration in 2 IF);
(2) total bandwidth (BW) of observation; 
(3) date of observation;
(4) time spent on the source excluding calibration overheads; 
(5) polarization;
(6) the flux calibrator used in the observation 
(flux density of the calibrator in Jy);
(7) the phase calibrator used during the observation 
(flux density of the calibrator in Jy); 
(8) half power beam width (HPBW\,$= 1.22 \cdot \lambda/D$; 
where $\lambda$ is the wavelength of observation and D is an antenna diameter); 
(9) synthesized beam achieved; 
(10) position angle (PA) of the restoring beam;
(11) rms error in the map;
(12) dynamic range achieved.
\end{minipage}
\end{table*}

The \citet{1977A&A....61...99Baars} absolute flux density scale was 
used to determine flux densities of the flux calibrators, 
phase calibrator, and the source. The calibrated data were 
channel collapsed after the bandpass calibration using the 
phase calibrator, and the deconvolved images were made using the 
task `IMAGR'. To account for wide-field imaging
with non-coplanar baselines, polyhedron imaging was used in `IMAGR'
(\citealt{1992A&A...261..353Cornwell}), where the field of view is subdivided
into a number of smaller fields (facets). These were $5\arcmin \times 5\arcmin$, 
$5\arcmin \times 5\arcmin$, and $7\arcmin \times 7\arcmin$\
at 629, 259 and 156\,MHz, respectively, facets covering the primary beam 
up to the half-power beam width (HPBW). Several rounds of {\it phase-based} 
self-calibration were performed iteratively,
by choosing point sources such that the flux density within
one synthesized beam is more than 5$\sigma$. Final maps were 
made from the full (u,v) coverage and the UV-data were weighted 
using Briggs robust weighting of 0 (\citealt{1995AAS...18711202Briggs}). 
The final images were then combined to reconstruct the sky using the task 
`FLATN' and the `FLATNed' image was then corrected for the primary 
beam\footnote{\texttt{http://ncra.tifr.res.in/ngk/primarybeam/beam.html}}
of the antenna using the task `PBCOR'.

The uncertainties in the estimated flux densities of the source (obtained using `TVSTAT' task) 
depend on the rms noise in the map as well as on the errors associated with
uncalibrated system temperature (T$_{sys}$) variations. These were taken to be 
5\% at 629 and 259\,MHz, and 8\% at 156\,MHz. The final error on the flux density 
was calculated using the expression:
\begin{equation}
\delta S = \sqrt{(S \times \sigma_{T_{sys}})^2 + \left (\sigma_{map} \times \sqrt{\frac{area}{beam}}  \right )^2}  
\end{equation}
\noindent
where S is the integrated flux density of the source, $\sigma_{T_{sys}}$ 
is the rms error due to the uncalibrated system temperature, $\sigma_{map}$
is the rms error in the map, $area$ is the area of source in pixels, and 
$beam$ is the beam area in pixels. 

The expected thermal noise values are about 0.1, 0.09 and 0.02 mJy/beam at 
156, 259 and 629\,MHz, respectively, for our observational set-up (Table\,\ref{obs}). 
However, the rms noise values achieved are about 20, 10 and 5 times 
worse, as a considerable amount of the {\emph {GMRT}} bad data was 
flagged during the reduction process. For the {\emph {VLA}} observations, the rms 
noise is close to the expected thermal noise $\sim$ 0.08 mJy/beam (Table\,\ref{obs}).

\section{Results}
\label{result}

The spectral index (defined here as $S_\nu \propto \nu^{\alpha}$) for the unresolved 
core of SHBL\,J001355.9$-$185406 has been evaluated using the peak flux 
values of the blazar obtained using the task `JMFIT' from the 156, 259 and 629\,MHz
{\emph {GMRT}} maps, along with the archival 352\,MHz data from the Westerbork 
In the Southern Hemisphere ({\emph {WISH}}) survey (\citealt{2002yCat.8069....0DeBreuck}), 
the 1400 MHz ({\emph {NVSS}} survey) data, and the {\emph {VLA}} data at 8434\,MHz.
By applying the weighted linear regression analysis we find $\alpha = -0.43 \pm 0.05$ 
(see Figure\,\ref{spec_radio}). The low-frequency radio spectrum seems to be even 
flatter than that, with $\alpha_{156}^{629} = - 0.1 \pm 0.1$ (see Table\,\ref{fluxden}),
as indeed expected for a BL Lac object (e.g., \citealt{1980ARA&A..18..321Angel}; 
\citealt{2004A&A...419..459Cavallotti}). 

\begin{figure}[t]
\includegraphics[height=8.0cm,width=8.0cm]{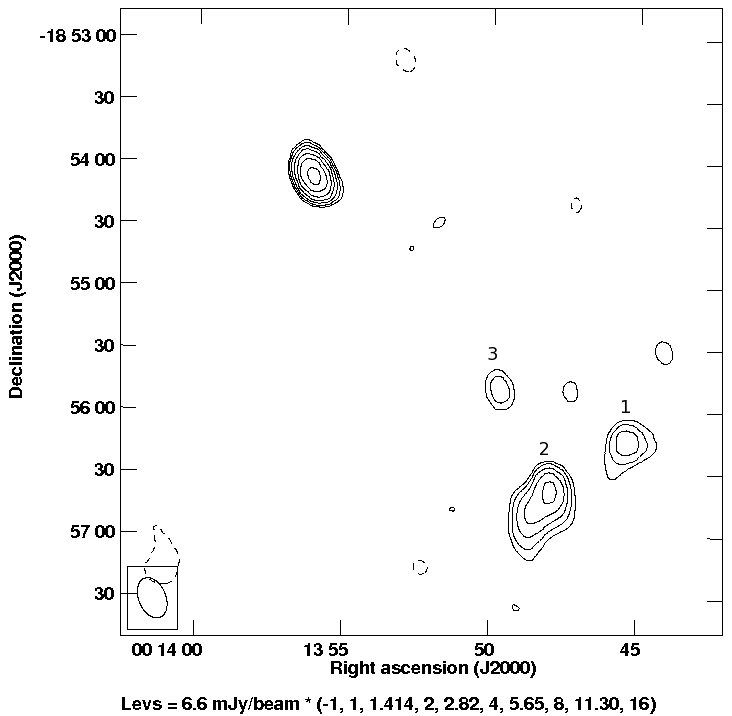}
\caption{156\,MHz {\emph {GMRT}} contour map. The bright point-like source
in the north-east corner of the image corresponds to the blazar core of SHBL\,J001355.9$-$185406. 
The newly resolved radio features to the south-west from the blazars are indicated with 
numbers 1, 2 and 3.}
\label{map.150}
\end{figure}

\begin{figure}[t]
\includegraphics[height=8.0cm,width=8.0cm]{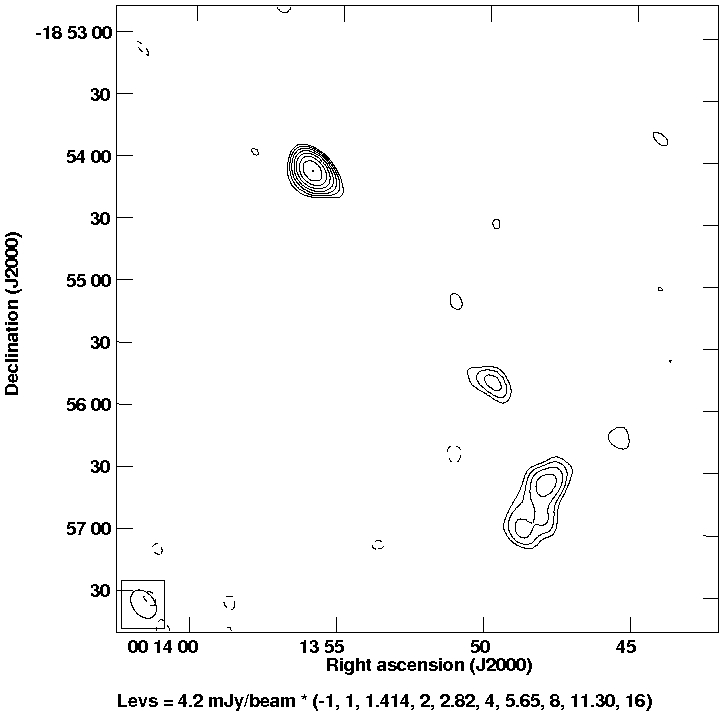}
\caption{259\,MHz {\emph {GMRT}} contour map. The bright point-like source
in the north-east corner of the image corresponds to the blazar core of SHBL\,J001355.9$-$185406.}
\label{map.240}
\end{figure}

\begin{figure*}
\begin{center}
\includegraphics[height=11cm,width=11.5cm]{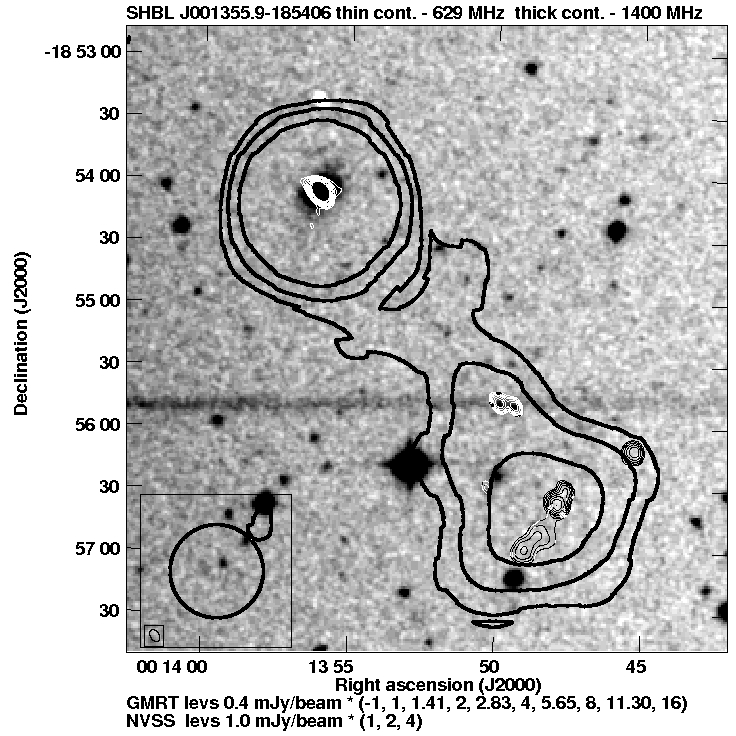}
\caption{629\,MHz {\emph {GMRT}} contour map (thin cont.). The bright point-like source
in the north-east corner of the image corresponds to the blazar core of SHBL\,J001355.9$-$185406.
The 1400\,MHz {\emph {NVSS}} map is represented by the thick contours, which are 
overlaid on the DSS $R$-band image. The bottom-left inset images shows 
the synthesized beams at 629 MHz (thin cont.) and 1400 MHz (thick cont.).}
\label{map.610}
\end{center}
\end{figure*}

The full-resolution {\emph {GMRT}} maps of the entire field of interest are
presented in Figure\,\ref{map.150}-\ref{map.610}. As shown, the lobe-like structure seen 
to the south-west from the blazar core in the {\emph {NVSS}} image breaks into three 
distinct features, marked as Source 1, Source 2 and Source 3 in Figure\,\ref{map.150}. 
Table\,\ref{fluxden} gives the integrated flux densities for the unresolved blazar core and the above 
mentioned sources. Detailed morphologies of all the detected objects are best 
displayed in the 629 MHz map, which is the best quality map out of the three obtained 
{\emph {GMRT}} maps (Table\,\ref{obs}); the corresponding Figure\,\ref{map.610} 
includes also the low-resolution 1400\,MHz contours for comparison, as well as
the overlaid {\emph {DSS}} $R$-band image. We note that at 8434\,MHz map, 
only the point-like blazar is detected, hence the corresponding image is not presented here.
 
\begin{figure}
\includegraphics[height=8.0cm,width=8.0cm]{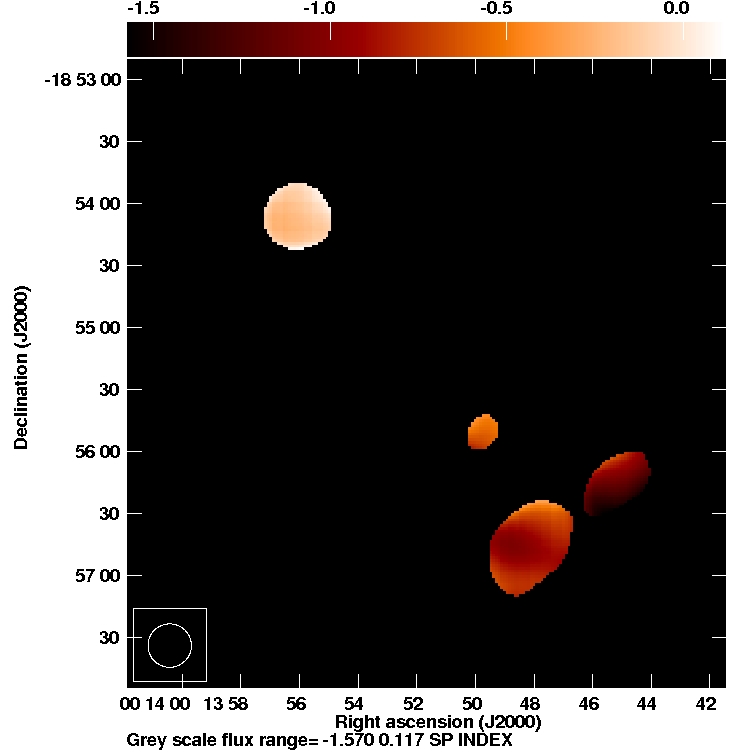}
\caption{The 156--629\,MHz spectral index distribution. Both 156 and 629\,MHz maps 
were convolved to a common circular beam of $21\arcsec \times 21\arcsec$.}
\label{spix}
\end{figure}

Figure\,\ref{spix} shows the spectral index map between 156 and 629\,MHz 
frequencies. When generating this map, the 629\,MHz images were made with 13$k\lambda$
uvrange (equal to the uvrange obtained at 156\,MHz with the {\emph {GMRT}}). This 
ensured the resultant synthesized beam in the 629\,MHz map close to that of 156\,MHz, 
(although not exactly equal due to the dependence on the filling of the {\it visibility} plane; 
e.g., \citealt{1999ASPC..180..151Cornwell}). Next the two maps were convolved to the 
common circular beam as large as the largest axis beam at 156\,MHz (i.e., 21\arcsec). 
As shown in the resulting figure, and already mentioned above, the unresolved blazar 
core displays a flat spectrum with $\alpha_{156}^{629} \simeq 0$; the three newly resolved
structures to the south-west are however characterized by steep spectra, 
$\alpha_{156}^{629} < - 0.5$ (see Table\,\ref{fluxden}).

\begin{table}
\tiny
\caption{Integrated radio flux densities obtained using `TVSTAT' in mJy and 156--629\,MHz spectral indices.}
\label{fluxden}
\begin{tabular}{cccccc}\\\hline
 Source       & 156\,MHz   & 259\,MHz & 629\,MHz & 8434\,MHz    & $\alpha_{156}^{629}$  \\
        & flux (err.) & flux (err.) & flux (err.) & flux (err.) &     \\ 
(1)     &  (2)        &(3)          &(4)          & (5)         & (6)  \\\hline
Blazar        & 55(6.0)      &  53(3.7)     &   45(2.2)    &  15(2.3)     &  $-$0.1(0.1)   \\
Source 1      & 29(5.3)      &  5 (1.6)     &   4.0(0.3)   & --           &  $-$1.4(0.2)  \\
Source 2      & 62(7.5)      &  47(4.0)     &   17(1.0)    & --           &  $-$0.9(0.1)   \\
Source 3      & 11(3.4)      &  15(2.4)     &   5.8(0.4)   & --           &  $-$0.5(0.3) \\ \hline
\end{tabular}

\begin{minipage}{0.48\textwidth}
Columns :(1) source name ; (2)--(5) integrated radio flux densities and the errors in parentheses 
at 156\,MHz, 259\,MHz, 629\,MHz and 8434\,MHz; (6) integrated spectral indices between
156--629\,MHz for the sources mentioned in the text.   
\end{minipage}
\end{table}

The {\emph {GMRT}} has a hybrid configuration, with 14 out of 30 antennas 
located in 1\,km square area (known as the central square).
The typical antenna spacing available in the central square with the {\emph {GMRT}}
is about $\sim 800$\,m, this assures sensitivity to sources 
with angular sizes up to $\sim 8\arcmin$, $\sim 5\arcmin$, and $\sim 2\farcm2$ 
at 156, 259 and 629\,MHz, respectively, even with the snap shot observations.
The half power beam 
width (HPBW), or, the primary beam at 156, 259 and 629\,MHz is
about 186\arcmin, 116\arcmin\, and 45\arcmin, respectively (Table\,\ref{obs}).
Since the extension of the lobe-like structure to the south-west from the 
blazar core is about 3\farcm3, the entire feature should be well within 
the primary beam at each frequency, and the availability of multiple 
short spacing in our observations ensures the measurements of the 
extended flux. Therefore, we do not expect any missing flux in our observations. 

We note, however, that the non-detection of the extended emission 
just near the blazar core as seen in the NVSS map (Figure \ref{fig.nvss}) could
be due 
to the limited sensitivity achieved in our observations (see, Table \ref{obs}). 
To investigate this issue in more detail, we computed
the 1400\,MHz expected flux density per 
5$^{\prime\prime}$ beam (as this is the typical synthesized beam 
achieved at 629\,MHz using the GMRT) near the blazar core.
Assuming a uniform flux density in the NVSS map in this region, 
we next evaluated the flux density
per beam at 629\,MHz 
assuming the spectral index $-1.0$. 
The extrapolated flux density per beam at 629\,MHz comes out to be 0.03 mJy, 
much below the rms error achieved in our 
observations. Hence, it is possible that the extended 
structure seen in the NVSS map just near the blazar 
core is real, but only below the sensitivity limit of 
our high-resolution maps at low frequencies.
Still, the radio morphologies of the sources detected to 
the south-west of blazar implies that the brightest part 
of the large-scale NVSS structure is unrelated to the blazar.
Source 1 can be easiliy classified as a 
ultra-steep spectrum source 
(e.g., \citealt{2000A&AS..143..303DeBreuck}, see Table \ref{fluxden}, Figure \ref{spix}) 
while Source 2 has a typical FR\,II radio morphology 
(e.g., \citealt{1980ARA&A..18..165Miley}; \citealt{2008ASPC..386...46Hardcastle}).
Source 3 breaks into two point-like sources, each with an optical 
counterpart (see Figure\,\ref{map.610}).
This strongly suggests that the detected features are related to 
distant radio galaxies, although the lack of any obvious 
optical counterparts (for Sources 1 and 2), or optical spectra (Source 3) 
precludes us from making any definite statements in this respect.

\section{Conclusions}
\label{conclutions}

In our study presented here, we investigated in detail large-scale 
radio morphology of the TeV-detected blazar SHBL\,J001356$-$185406 which, 
on the low-resolution {\emph {NVSS}} map, appears to be of the FR\,II type. 
High-sensitivity and high-dynamic range (Table\,\ref{obs}) 
observations (which do not suffer from a missing flux; see Section\,\ref{result}) 
were made using the {\emph {GMRT}} at low frequencies of
156, 259 and 629\,MHz. 
Our analysis indicated that: (1) the integrated radio spectral index is 
flat down to 156\,MHz for the blazar core (Figure\,\ref{spec_radio}); 
(2) no diffuse arcmin-scale emission is present around the unresolved 
blazar core (see Figures\,\ref{map.150}--\ref{map.610}); 
(3) the 3\farcm3-long, lobe/jet-like structure extending to the south-west
from the blazar core on the {\emph {NVSS}} map breaks into three distinct,
steep-spectrum radio features, unrelated to the blazar but instead associated
most likely with background AGN.

As a corollary to this study we point out that the observations presented 
here are rather unique, since very low-frequency, high-resolution and high-dynamic 
range studies of TeV detected HBLs are particularly sparse. The new data allow us to
constrain the upper limits for the extended radio halo around the blazar
as $< 12\%$ at 156\,MHz, $< 8\%$ at 259\,MHz, and $< 0.9\%$ at 629\,MHz 
of the core emission. This is agreement with the early {\emph {VLA}} studies of
large-scale structures of BL Lacs at GHz frequencies by 
\citet{1984AJ.....89..189Ulvestad} and \citet{1986AJ.....92....6Ulvestad}, who
found that only a few of the objects from the analyzed samples possess any
arcminute-scale structures with total fluxes exceeding $\sim 0.1\%$ of the core
emission.

The another interesting result of our study is that the radio 
spectral index of the SHBL\,J001356$-$185406 core remains flat down 
to 156\,MHz, with no dramatic curvature. Flat spectra (indices $\alpha \sim 0$)
of relativistic jets observed at small viewing angles are widely believed to
arise due to a superposition of self-absorbed spectra of different parts
of an outflow with conserved magnetic flux and maintained
power-law spectrum of the emitting electrons \citep{1979ApJ...232...34Blandford}.
The fact that in the case of the SHBL\,J001356$-$185406 core we do not 
see any spectral turnover down to 156\,MHz indicates therefore that a
self-similar character of the jet in the source holds up to relatively large 
distances from the jet base.

\begin{acknowledgements} 
We thank the anonymous referee for the helpful suggestions.
We thank the staff of the {\emph {GMRT}} that made these observations 
possible. {\emph {GMRT}} is run by the National Centre for Radio
Astrophysics of the Tata Institute of Fundamental Research.
This research has made use of NASA/IPAC Extragalactic Database 
(NED), which is operated by the Jet Propulsion 
Laboratory, California Institute of Technology, under contract with National 
Aeronautics and Space Administration. The National Radio Astronomy 
Observatory is a facility of the National Science Foundation 
operated under cooperative agreement by Associated Universities, Inc. 
This work was supported by the Polish National Science Centre through 
the grant DEC-2012/04/A/ST9/00083. M.J. is supported by Polish National 
Science Center grant  DEC-2013/09/B/ST9/00599.
\end{acknowledgements} 
\clearpage


\end{document}